\documentclass[pra,twocolumn,showpacs,preprintnumbers,superscriptaddress]
{revtex4-1}

\usepackage{times}
\usepackage{bm}
\usepackage{graphicx}
\usepackage{amsbsy}
\usepackage{amsmath}
\usepackage{amsfonts}
\usepackage{amsthm}
\usepackage{color}

\begin{document}

\theoremstyle{plain}
\newtheorem{theorem}{Theorem}
\newtheorem{lemma}[theorem]{Lemma}
\newtheorem{corollary}[theorem]{Corollary}
\newtheorem{proposition}[theorem]{Proposition}
\newtheorem{conjecture}[theorem]{Conjecture}

\theoremstyle{definition}
\newtheorem{definition}[theorem]{Definition}

\theoremstyle{remark}
\newtheorem*{remark}{Remark}
\newtheorem{example}{Example}

\title{More Communication with Less Entanglement}

\author{Pankaj Agrawal}
\thanks{agrawal@iopb.res.in}
\affiliation{Institute of Physics, Sainik School Post,
Bhubaneswar-751005, Orissa, India}
\author{Satyabrata Adhikari}
\thanks{tapisatya@gmail.com}
\affiliation{Institute of Physics, Sainik School Post,
Bhubaneswar-751005, Orissa, India}
\author{Sumit Nandi}
\thanks{sumit@iopb.res.in}
\affiliation{Institute of Physics, Sainik School Post,
Bhubaneswar-751005, Orissa, India}

\date{\today}

\begin{abstract}

We exhibit the intriguing phenomena of ``Less is More'' using a
set of multipartite entangled states. We consider the quantum
communication protocols for the {\em exact} teleportation,
superdense coding, and quantum key distribution. We find that
sometimes {\em less} entanglement is {\em more} useful. To
understand this phenomena we obtain a condition that a resource
state must satisfy to communicate a $n$-qubit pure state which has $m$
terms. We find that an appropriate partition of the resource state
should have a von-Neumann entropy of ${\rm log}_{2} m$.
Furthermore, it is shown that some states may be suitable for
{\em exact} superdense coding, but not for {\em exact} teleportation.

\end{abstract}


\maketitle

\newcommand{\bra}[1]{\langle #1|}
\newcommand{\ket}[1]{|#1\rangle}
\newcommand{\braket}[2]{\langle #1|#2\rangle}

\section{Introduction}

The entanglement has been used as a quantum resource to carry out
a number of communication tasks such as teleportation
\cite{bbcjpw}, superdense coding \cite{bw}, telecloning
\cite{plentel}, secret sharing \cite{hillery}, quantum-key distribution (QKD)
\cite{bennett,ekert,gisin}. The entangled resource state may
have bipartite or multipartite entanglement. A number of protocols
which were first introduced in the context of a bipartite system
can be extended to a multipartite system. However, in the case of
multipartite systems, the nature of entanglement is still not
fully understood  \cite{horodecki4}. Also, there can be many
variants and extensions of such protocols in this new setting.
The multipartite entangled states can be classified according to
various schemes  \cite{gour,sharma2,li}. Different classes exhibit
different types of entanglement properties. In this letter,
we examine a set of four-qubit states that belong to different
categories according to SLOCC \cite{vddv} classification.
We use these states to explore some of the variations
of the bipartite protocols. Such
studies may even allow a better understanding of multiparticle
entanglement and classification of quantum states according to
their ability to carry out a specific task. SLOCC classification
is not a useful guide for the suitability of a state to be
a resource state for a specific task \cite{ap}.

We are interested in the {\em exact} teleportation, superdense
coding and quantum key distribution. These tasks may be carried
out nonmaximally. A multiqubit state shared by two parties, allows
many possibilities. A resource state is suitable for {\em exact}
teleportation if it can be used to teleport a $n$-qubit state with
$m$ terms, for some $m$ and $n$, with unit probability and unit
fidelity. By {\em exact} superdense coding we mean the ability to
communicate $n+1$ or higher integral values of classical bits by
transmitting $n$ qubits.  For the case of quantum key
distribution, one should be able to generate a key using the
resource state. We use
 von Neumann entropy of the subsystems as a tool for
 our investigations. We note that
one would need a task-oriented maximally entangled
state (TMES) \cite{apr} to carry out a task maximally. A prescription
was given in \cite{apr} to construct such TMESs.

In this paper, we consider a set of four-qubit states and show that
larger entanglement is not always more useful for some communication 
tasks. There has been some discussion of this issue in the literature
\cite{eberherd,acin}.
Our investigations would lead us to ask the question: Given a resource
state, what unknown states can be teleported exactly ? Or, given a
state to be teleported, what are the minimum requirement for
the resource state ?


 We consider the following
inequivalent quadripartite entangled states of qubits in terms of SLOCC \cite{vddv},
\begin{eqnarray}
 \ket{GHZ}& = & \frac{1}{\sqrt{2}}(\ket{0000}+\ket{1111}), \\
 \ket{W}& = &\frac{1}{2}(\ket{0001}+\ket{0010}+\ket{0100}+\ket{1000}),\\
 \ket{\Omega}& = &\frac{1}{2}(\ket{0000}+\ket{0110} +
\ket{1001} - \ket{1111}),  \\
 \ket{S_{1}} &  = & \frac{1}{2}(\ket{0000}+\ket{0101}+\ket{1000}+\ket{1110}),  \\
 \ket{S_{2}} &  = & \frac{1}{2}(\ket{0000}+\ket{1011}+\ket{1101}+\ket{1110}).
\end{eqnarray}

 These states have genuine quadripartite entanglement in the sense that they
 cannot be written as product states in any partition.
 Among these five entangled states,
only $\ket{GHZ}$ and $\ket{W}$ states are
symmetric with respect to the permutation of qubits; thus any quantum
information task performed using these states is independent of distribution
of particles among the parties.
The $\ket{\Omega}$  state is the cluster state introduced by Briegel and
 Raussendorf  \cite{br}.
This state has been discussed extensively in the context of one-way
quantum computation.

It will be useful to catalog the von Neumann entropy (henceforth,
called entropy) of all the bipartite partitions of theses states. 
The entropy for a
state $\rho$ is defined as $S(\rho) = - \rm{Tr}(\rho
\rm{log_{2}}(\rho))$. We have four-qubit states. We can label these qubits
as ``$1,2,3,4$''. Then $\rho_1$ is the reduced density matrix of
the qubit with label `1';  $\rho_{12}$ is the reduced density matrix of
the qubits with labels `1' and `2' and so on. In a bipartite partition, 
both subsystems of four-qubits
will have identical entropies. For example, $S(\rho_1) =
S(\rho_{234})$, $S(\rho_{12}) =  S(\rho_{34})$ etc. This can be seen by considering
the Schmidt decomposition of the states.

In Table I, we list the entropies of the subsystems for the above five states.

\begin{center}
\begin{tabular}{|c|c|c|c|c|c|c|c|c|}
\hline
 States &$S(\rho_1)$&$S(\rho_2)$&$S(\rho_3)$&$S(\rho_4)$&$S(\rho_{12})$&$S(\rho_{13})$
&$S(\rho_{14})$\\
\hline
GHZ&1&1&1&1&1&1&1\\
\hline
$\Omega$  &1&1&1&1&2&2&1\\
\hline
W&0.81&0.81&0.81&0.81&1&1&1\\
\hline
$S_1$&0.81&1&0.81&0.81&1.5&1.22&1.22\\
\hline
$S_2$&0.81&1&1&1&1.5&1.5&1.5\\
\hline
\end{tabular}

\vspace{0.2in}

 {Table I: Entropies of the subsystems}
\end{center}

When the four-qubit system is partitioned into `1' and
`234' parts, then each subsystem will have entropy $S(\rho_1)$;
for the partition `14' and `23', each subsystem
will have entropy  $S(\rho_{14})$. Other entries of the table
can be understood in the similar manner.
In our analysis of above four-qubit states, these
entropies will play an important role. In fact,
the success of various protocols
depends on the values of these entropies.

\section{Teleportation and Entropy}

   With a four-qubit resource state, one can look for teleporting
   one-qubit, two-qubit, or three-qubit unknown states with varying
   number of terms. As we shall see, usefulness of a resource
   state would depend on the
   entropies of its subsystems. For maximal teleportation, a subsystem
   must have maximal entropy. For non-maximal teleportation, smaller
   entanglement is enough.


 Let us consider first when Alice wishes to teleport an
unknown qubit state $\ket{\psi}_{a} = \alpha \ket{0}_{a} + \beta
\ket{1}_{a}$ to Bob.
 The qubits of the resource states are distributed in
such a way that Bob will have one qubit and Alice the rest.
To find the usefulness of a resource state, one can look at the
entropies of the subsystems. For any type of teleportation to
succeed, the entropy of the Bob's qubit needs to be one. With the
$\ket{GHZ}$ state the teleportation protocol will work for any
distribution of the particles because all individual qubits of a
$\ket{GHZ}$ state have unit entropy. The $\ket{\Omega}$ state is
not symmetric under the permutations of qubits. Still, the
teleportation would succeed regardless of which qubit Bob has
because all individual qubits have entropy as one. 
 For example, let Alice has the qubits 1, 2, 3 and Bob has the
qubit 4. The combined state of the the five qubits can be
rewritten as
\begin{eqnarray}
\ket{\psi}_a \ket{\Omega}_{1234}
& =  & [\ket{\Omega_1^+}_{a123}
\sigma_0 \ket{\psi}_{4} +\ket{\Omega_1^-}_{a123} \sigma_3 \ket{\psi}_{4}+ \nonumber \\
& & \ket{\Omega_2^+}_{a123}\sigma_1 \ket{\psi}_{4}
+\ket{\Omega_2^-}_{a123} i\sigma_2 \ket{\psi}_{4}]/2, \nonumber
\end{eqnarray}
 where
$\ket{\Omega_1^{\pm}} =
(\ket{00}\ket{\varphi^+}\pm\ket{11}\ket{\varphi^-})/\sqrt{2}$,
$\ket{\Omega_2^{\pm}} =
(\ket{01}\ket{\varphi^-}\pm\ket{10}\ket{\varphi^+})/\sqrt{2}$ and
$\ket{\varphi^\pm } =  (\ket{00}\pm\ket{11})/\sqrt{2}$. With these
measurement vectors, the teleportation protocol can be carried out
with two cbit of classical communication \cite{pba}.

  One cannot
teleport the state $\ket{\psi}_{a}$ using $\ket{W}$ state as a
 resource state since none of the qubits has entropy as one. However, the
$\ket{W_{mn}}$ state of \cite{ap,pba} can be used for the teleportation.
The $\ket{S_1}$ and $\ket{S_2}$ states are not symmetric under
the permutations of the qubits; therefore only specific distributions
of the qubits leads to the successful teleportation. For the
$\ket{S_1}$ state, only when Bob has the qubit 2, the protocol works.
This is because only second qubit has entropy as one.
 In the case of $\ket{S_2}$ state Bob can have any qubit, except
the qubit 1.

We now consider the possibility of
teleporting an unknown arbitrary two-qubit state $ \ket{\psi}_{ab}
= \alpha \ket{00}_{ab}+\beta \ket{01}_{ab}+\gamma \ket{10}_{ab}+
\delta \ket{11}_{ab}$.
It would be possible with only a few four-qubit entangled
states. For any type of teleportation of an arbitrary and unknown
two-qubit state to succeed, the entropy of the Bob's two qubits
should be two. For the teleportation of the restricted subclass,
with two arbitrary parameters, the entropy needs to be one only.
In such a situation, one can find suitable measurement vectors.
The $\ket{\Omega}$ state
be used to teleport an arbitrary two-qubit unknown state for all
appropriate distributions of qubits, except the partition of (1,4)
and (2,3) qubits. We can see the protocol by rewriting the
six-qubit state as \cite{pba}
\begin{eqnarray}
\ket{\psi}_{ab}\ket{\Omega}_{1234}& =& \nonumber \\
& & \hspace{-.7in} \sum_{i=1}^{2}  [\ket{\Omega^i_+}_{ab12}\;U^{i}_+\;\ket{\psi}_{34} +
 \ket{\Omega^i_-}_{ab12}\;U^{i}_-\;\ket{\psi}_{34} +  \nonumber \\
&  & \hspace{-.7in} \ket{\Omega^{i+2}_+}_{ab12} U^{i+2}_+\ket{\psi}_{34}
+\ket{\Omega^{i+2}_-}_{ab12} U^{i+2}_-\ket{\psi}_{34}+  \nonumber \\
& & \hspace{-.7in}  \ket{\Omega^{i+4}_+}_{ab12}  U^{i+4}_+ \ket{\psi}_{34} +
\ket{\Omega^{i+4}_-}_{ab12} U^{i+4}_- \ket{\psi}_{34}+ \nonumber \\
& & \hspace{-.7in} \ket{\Omega^{i+6}_+}_{ab12} U^{i+6}_+ \ket{\psi}_{34}
+\ket{\Omega^{i+6}_-}_{ab12} U^{i+6s}_-\ket{\psi}_{34}]/4,
\end{eqnarray}
where,
\begin{eqnarray}
\ket{\Omega^i_{\pm}}  & = & (\ket{0}\ket{\eta^+_i}\ket{0} \pm  \ket{1}\ket{\eta^-_i}\ket{1})/\sqrt{2}    \nonumber \\
\ket{\Omega^{i+2}_{\pm}} & =  & (\ket{0}\ket{\eta^-_i}\ket{0} \pm  \ket{1}\ket{\eta^+_i}\ket{1})/\sqrt{2}  \nonumber \\
\ket{\Omega^{i+4}_{\pm}} & = & (\ket{0}\ket{\eta^+_i}\ket{1} \pm  \ket{1}\ket{\eta^-_i}\ket{0})/\sqrt{2}     \nonumber\\
\ket{\Omega^{i+6}_{\pm}} & = & (\ket{0}\ket{\eta^-_i}\ket{1} \pm  \ket{1}\ket{\eta^+_i}\ket{0})/\sqrt{2}
\end{eqnarray}
Here $i = 1,2$ and $\ket{\eta^\pm_1} = \ket{\varphi^\pm} =
\frac{1}{\sqrt{2}}(\ket{00}\pm\ket{11})$, and $\ket{\eta^\pm_2} = \ket{\psi^\pm} =
\frac{1}{\sqrt{2}}(\ket{01}\pm\ket{10})$. $U^{i}_\pm$ are appropriate
unitary operators.

The GHZ state is
not a suitable resource state for the teleportation of an
arbitrary unknown two-qubit state. However, an entangled two-qubit
state $\ket{\psi_{1}}_{ab} = \sigma_i\sigma_j(\alpha \ket{00}_{ab}
+ \beta\ket{11}_{ab})$ can be teleported.
 One can teleport the state because the entropies of two-qubit subsystems
of the GHZ-state is one. This state superficially looks like
one-qubit state.
Similarly, in the case of the $\ket{W}$-state
 the two-qubit subsystems have entropy one, so one can
teleport some subclasses of arbitrary two-qubit states.
For example,  one can teleport the state
 $\ket{\psi_{2}}_{ab} = \alpha \ket{00} + \beta \ket{\psi^+}.$
In the case of $\ket{S_1}$ and $\ket{S_2}$ states, two-qubit
subsystems have entropy more than one but less than two.
Surprisingly, although two-qubit subsystems have larger entropy
than in the case of W-state or GHZ-state, but cannot teleport even
the subclasses of states. Therefore, {\em a larger entanglement
does not necessarily help in teleportation}.

 For the teleportation of a general three-qubit state,
one would need an entangled state of six qubits \cite{pba}. From
the Table I, we see that the maximum entropy of three-qubit
subsystems for the states under consideration is one. Therefore
one can teleport at most a state with two terms only. For example,
although one can use $\ket{\Omega}$ state to teleport a two-qubit
state with four unknown parameters, but it can teleport a three-qubit
state with only two unknown parameters.


\section{Condition for Teleportation }

We saw in the last section that the entropy of the
subsystems of the given resource entangled state plays
an important role. Now we
ask the question that if we are given a $n$-qubit state with $m$ terms
to teleport, what kind of resource state is needed? In a reverse
way, the question can be posed as: given a resource state, what
are the states that can be teleported using this resource? The
answer to this question tells us that why sometimes a more
entangled state is less suitable.

\textbf{Theorem:} A resource state is useful to teleport an
unknown $n$-qubit state which has $m$ terms
if and only if the resource states qubits could be
distributed in such a way that the receiver's $n$ qubits have
entropy ${\rm log}_{2} m$.

 \textbf{Proof:} Let us consider an unknown $n$-qubit state with $m$-terms that
  Alice wishes to teleport to Bob

\begin{equation}
  \ket{\Psi}_{n} =   \sum_{k=1}^{m} \alpha_{k} \ket{\eta_{k}}_{n}.
\end{equation}

 The state is normalized and the basis set is orthonormal
\begin{equation}
     \braket{\eta_k}{\eta_l} = \delta_{kl}, \;\;\;\;\;\;\;\;\;\;
   \sum_{k=1}^{m} |\alpha_{k}|^2 = 1.
\end{equation}

   Let the resource state be a $N$-qubit state. For this
 resource state to be useful, we should be able to write
 it as,
\begin{equation}
  \ket{R}_{N} =   {1 \over \sqrt{m}} \sum_{l=1}^{m} \ket{\chi_{l}}_{N-n} \ket{\eta_{l}}_{n}.
\end{equation}

 Here the states $\ket{\chi_{l}}_{N-n}$ may not be orthonormal. The combined state can be
written as

\begin{eqnarray}
       \ket{\Psi}_{n} \ket{R}_{N} & = &
    {1 \over \sqrt{m}} \sum_{k=1}^{m}  \sum_{i=1}^{m}  \alpha_{i} \ket{\eta_{i}}_{n}  \ket{\chi_{k}}_{N-n} \ket{\eta_{k}}_{n} \nonumber \\
      & =  &
    {1 \over \sqrt{m}} \sum_{k=1}^{m}  \sum_{i=1}^{m}  \ket{\eta_{i}}_{n}  \ket{\chi_{k}}_{N-n} \alpha_{i} \ket{\eta_{k}}_{n}.
\end{eqnarray}

  Alice will now make a measurement in an orthonormal basis $ \ket{\theta_{l}}_{p}$.
Therefore, we should be able to write

\begin{equation}
   \ket{\eta_{i}}_{n}  \ket{\chi_{k}}_{N-n} =   {1 \over \sqrt{m}} \sum_{l=1}^{m^2}  C_{ik,l}
                    \ket{\theta_{l}}_{N}. \label{measurementbasis}
\end{equation}

  The $C_{ik,l}$ is an interesting object. For each $l$, it is a
  $ m \times m$ matrix in `$ik$' space.
 It is also a $m^2 \times m^2$ matrix with row label as $ik$. We need to find
 a condition on $C_{ik,l}$ such that we indeed have a suitable measurement
basis. Using the above equation

\begin{eqnarray}
       \ket{\Psi}_{n} \ket{R}_{N} & = &
    {1 \over m} \sum_{i=1}^{m}  \sum_{k=1}^{m}  \sum_{l=1}^{m^2} C_{ik,l}
                    \ket{\theta_{l}}_{N} \alpha_{i} \ket{\eta_{k}}_{n}  \nonumber \\
      & =  &
    {1 \over m} \sum_{l=1}^{m^2} \ket{\theta_{l}}_{N}  \sum_{k=1}^{m}  \sum_{i=1}^{m} C_{ik,l}
                    \alpha_{i} \ket{\eta_{k}}_{n}.  \nonumber
\end{eqnarray}

 For the teleportation to succeed, we should have,
\begin{equation}
\sum_{k=1}^{m}  \sum_{i=1}^{m} C_{ik,l} \alpha_{i}
\ket{\eta_{k}}_{n} = V^{l}  \sum_{n=1}^{m} \alpha_{n}
\ket{\eta_{n}}_{n}.
\end{equation}

 Taking its adjoint and scalar product with itself, we get,
\begin{equation}
      \sum_{k=1}^{m}  \sum_{i=1}^{m} \sum_{i^\prime=1}^{m} C_{ik,l} C^{*}_{i^\prime k,l} \alpha_{i}
             \alpha^{*}_{i^\prime} = 1.
\end{equation}

 This condition is true for each $l$,
\begin{equation}
      \sum_{i=1}^{m} \sum_{i^\prime=1}^{m} (C C{^\dagger})_{ii^\prime,l} \alpha_{i}
             \alpha_{i^\prime} = 1.
\end{equation}

 For this equation to be satisfied, we must have
\begin{equation}
 (C C{^\dagger})_{ii^\prime,l} = \delta_{ii^\prime}.
\end{equation}

 This suggests that $C$ is unitary in `$ik$' space for
each $l$ for teleportation to succeed. Let us now see what it
means for the resource state.

Taking the adjoint and the scalar product,
\begin{equation}
    \braket{\eta_i^\prime}{\eta_i} \braket{\chi_k^\prime}{\chi_k} =
    {1 \over m} \sum_{l=1}^{m^2} \sum_{l^\prime=1}^{m^2} C_{ik,l} C^{*}_{i^\prime k^\prime,l^\prime} \braket{\theta_l}{\theta_l^\prime},
\end{equation}

\begin{equation}
    \delta_{i i^\prime} \braket{\chi_k^\prime}{\chi_k} =
    {1 \over m} \sum_{l=1}^{m^2} C_{ik,l} C^{*}_{i^\prime k^\prime,l}.
\end{equation}

 Multiplying by $\delta_{i i^\prime}$ and summing over
$i^\prime$ and $i$, we get,
\begin{equation}
    \braket{\chi_k^\prime}{\chi_k} =
    {1 \over m^2} \sum_{l=1}^{m^2} (C^\dagger C)_{kk^\prime,l}.
\end{equation}

Since $C$ is unitary in the subspace, we get
\begin{equation}
    \braket{\chi_k^\prime}{\chi_k} = \delta_{kk^\prime}.
\end{equation}

 Therefore $\ket{\chi_{k}}$ should be orthonormal for the
exact teleportation. Thus
the resource state should have the form

\begin{equation}
  \ket{R}_{N} =   {1 \over \sqrt{m}} \sum_{l=1}^{m} \ket{\chi_{l}}_{N-n} \ket{\eta_{l}}_{n},
\end{equation}
 with both $\ket{\chi_{k}}$ and $\ket{\eta_{l}}$ being
orthonormal. So, the entropy of the reduced density matrix of the
Bob's qubits would be $\rm{log}_{2}m$.

Conversely, if the entropy of the reduced density matrix of the
Bob's qubits is $\rm{log}_{2}m$ then we can always find a
measurement basis given in (\ref{measurementbasis}) so that one
can do faithful teleportation of $n$-qubit state with $m$-terms.

 What we have shown is that if we wish to teleport a $n$-qubit
  state with $m$ terms, then we should be able to distribute resource
  states qubits in such a way such that Bob's $n$ qubits have entropy
  ${\rm log}_{2} m$. Given a resource state, we can compute entropy of all the
partitions. If there is a partition where Bob's $n$
qubits have entropy as ${\rm log}_{2} m$, then the state with
$m$ terms can be teleported with that partition.

\section{Superdense coding and Entropy}

We now discuss the superdense coding capacity of various entangled
states under discussion. There are three possible scenarios.
In the scenario 1 (SN1), Alice has one qubit of the resource
state; in the scenario 2 (SN2), Alice has two qubits of the
resource state; in the scenario 3 (SN3), Alice has three qubits of
the resource state. In each case, Bob has the rest of the qubits
of the resource state.
For all such distributions, they follow the standard superdense
coding protocol to transmit a classical message. In this protocol,
Alice applies unitary operations $ \sigma_0, \sigma_1, i\sigma_2,
\sigma_3$ with equal probabilities on her qubits and sends them to
Bob. Bob performs a joint measurement on all the four qubits to
retrieve the original message. Since some states are asymmetric
with respect to the permutation of qubits, the distribution may
affect the superdense coding capacity of the states. The capacity
mainly depends on how many orthogonal states are obtained using
unitary transformations by Alice. This is because orthogonal
states can be perfectly distinguished. Therefore, the task is to
find out the number  of orthogonal states that can be obtained by
unitary operations on the particles possessed by the sender.

The  superdense coding capacity also appears to depend on the entropy
of the  subsystem on which Alice applies the unitary operations.
This is specially true for maximal superdense coding. It also seems
to be the case for non-maximal superdense coding, as we will see below.
 The $\ket{\Omega}$ state is
the best quantum resource from the point of view of superdense
coding.
 The classical capacity in the $SN1$ scenario is two classical bits
(cbits) irrespective of distribution of the qubits, as the entropy
of each individual qubit is one.
In scenario $SN2$, Alice can transmit four cbits by transmitting
two qubits with specific distributions of qubits. The entropy for
the subsystems (1,2) or (1,3) is two, while it is one for the
subsystem (1,4). Therefore, the classical capacity is 4 cbits when
Alice has subsystems (1,2) (or (3,4)) or (1,3) (or (2,4)), but it
is three cbits when Alice has the subsystem (1,4) (or (2,3)). To
illustrate this let us consider the case when Alice has particles
1 and 2. On applying unitary transformations, $\sigma_k \otimes
\sigma_\ell \otimes \sigma_0\otimes \sigma_0$ $(k, \, \ell =
0,1,2,3)$,  the sixteen orthogonal states are obtained. Therefore,
clearly the classical capacity is four cbits which is the maximum
possible.

The $\ket{GHZ}$ state is
symmetric under the permutation of particles. In the
scenario SN1, Alice can
produce four orthogonal states, so she can transmit 2 cbits of
information. This is possible because the entropy of the
Alice's qubit is one. In the scenario SN2, Alice applies
unitary operations on her two qubits, giving only eight
orthogonal states.  It allows Bob to
access only three cbits. In the scenario SN3, Alice can
transmit four cbits of information.
In general using $n$-qubit GHZ-state one may be able to send $n$
bits of classical information by sending $n-1$ qubits. We note
that this state is not suitable for maximal superdense coding.
 For the W-state, in the scenario $SN1$ and $SN3$, this state is
 not suitable for superdense
 coding  because the entropy of Alice's subsystem is less than
 one. However a variant of this state,
 $\ket{W_{mn}}$, \cite{ap,pba} can be used.
 In $SN2$ scenario, Bob can access three cbits,
because the two-qubit subsystems have the entropy as one.

For the $\ket{S_1}$ state the success of the protocol depends
on the distribution of the
particles between the parties. In the $SN1$ scenario, the protocol
succeeds only when Alice has the qubit 2. This is because the
entropy of this qubit is one. Other qubits have individual entropy
0.81 which is less than 1.
In the scenario $SN2$, if Alice has qubits (1,2) and she
applies unitary operations on her qubits, she gets eight orthogonal
states. Therefore the capacity with this distribution is three cbits.
However, if Alice has qubits (1,3) or (1,4), then unitary
transformations yield at most four orthogonal states. So there is
no enhancement in the classical capacity. We note that the entropy
of the subsystem (1,2) is 1.5, while it is 1.22 for the subsystems
(1,3) and (1,4). Here,  we observe that there is an enhancement in
the capacity for the entropy 1 (as earlier) and 1.5, but not for
all values greater than 1. Surprisingly, the capacity of
superdense coding is two cbits in spite of the entropy of two
qubit subsystem being greater than one. It should be noted a
similar observation is made in case of entangled qudits \cite
{mozes}.
In the $SN3$ scenario, Alice will be able to transmit four cbits
by sending three qubits only when she has qubits 1, 3 and 4. This
is, as earlier, because only qubit 2 has entropy as one.
We observe that when Alice has qubits (1,2), she can do superdense
coding, but she would not be able to teleport a state. So we have
a situation where {\em a state is suitable for exact superdense coding
but not for exact teleportation}.
For the $\ket{S_2}$ state, the distribution of the qubits is also
important. The protocol can be implemented in the $S{1}$ scenario
for all possible distribution of qubits except when Alice has the
qubit 1. In the SN2 scenario, like
$\ket{S_1}$ state, we can use the state for superdense coding
for all distributions. This is because now all such partitions
have entropy 1.5. The channel capacity, in each case is 3 cbits.
It appears that after 1, the entropy value of 1.5 has special
significance.
This state is not suitable for exact teleportation, but can
be used for non-maximal superdense coding.

\section{Quantum key distribution}

In this section, we consider the quantum
information processing task of quantum key distribution (QKD)
and show that it works well with lesser amount of shared entanglement.
Let us consider 2 four-qubit states --
 $|GHZ\rangle$ and $|S_{1}\rangle$, which were defined earlier.
Let us suppose that Alice and Bob possess two qubits each. We see
from Table I that the two-qubit subsystems of $|S_{1}\rangle$
state have more entanglement than the two-qubit subsystem of the
$|GHZ\rangle$ state.
However, we will see that $|S_{1}\rangle$ state cannot be used in
a variant of BB84 QKD scheme \cite{bennett} while $|GHZ\rangle$
state can be used.
Let us consider $|GHZ\rangle$ state, given in (1) . Here Alice
holds the qubits 1 and 3 and the remaining two qubits 2 and 4 are
with Bob. In the Bell basis
$\{|\Phi^{+}\rangle,|\Phi^{-}\rangle,|\Psi^{+}\rangle,|\Psi^{-}\rangle\}$,
$|GHZ\rangle$ state can be re-written as
\begin{eqnarray}
|GHZ\rangle_{1234}=\frac{1}{\sqrt{2}}(|\Phi^{+}\rangle_{13}\,|\Phi^{+}\rangle_{24}
+|\Phi^{-}\rangle_{13}\,|\Phi^{-}\rangle_{24}).
\label{ghzbellbasis}
\end{eqnarray}

Here, $|\Phi^{\pm}\rangle = {1 \over \sqrt{2}} (|00\rangle \pm |11\rangle)$ and
 $|\Psi^{\pm}\rangle = {1 \over \sqrt{2}} (|01\rangle \pm |10\rangle)$.
In the next step, Alice randomly performs measurements on her
particles in either $\{|00\rangle,|11\rangle\}$ basis or in the
basis $\{|\Phi^{+}\rangle,|\Phi^{-}\rangle\}$. The information is
encoded by using the two binary digits $0$ and $1$. If the
measurement outcome is $|00\rangle_{13}~
(|\Phi^{+}\rangle_{13})$ then $0$ is encoded and if the
measurement outcome is $|11\rangle_{13}~
(|\Phi^{-}\rangle_{13})$ then $1$ is encoded. After Alice's
measurement, Bob also randomly chooses either the basis
$\{|00\rangle,|11\rangle\}$ or the basis
$\{|\Phi^{+}\rangle,|\Phi^{-}\rangle\}$ and then performs
measurement on his particles in that basis. In the next stage,
Alice publicly announces the basis in which she had measured the state
of the particles but does not declare the measurement outcome.
If Bob finds that his measurement basis
matches with the Alice's basis, then he informs Alice to keep the data,
otherwise the data is thrown out. In this way, quantum key
can be distributed between Alice and Bob. Therefore,
$|GHZ\rangle$ state can be used in generating the quantum
key.

Now the question arises whether the state
$|S_{1}\rangle$ is also suitable for this QKD scheme? To answer this
question, we have to write the four-qubit state $|S_{1}\rangle$ in
the computational basis as well as in the Bell basis.
In the Bell basis, this state can be re-expressed as
\begin{eqnarray}
|S_{1}\rangle_{1234}&=&\frac{1}{4}[(2|\Phi^{+}\rangle_{13}+2|\Phi^{-}\rangle_{13}
+|\Psi^{+}\rangle_{13} +|\Psi^{-}\rangle_{13})\nonumber\\&\,&
|\Phi^{+}\rangle_{24}+
(|\Psi^{+}\rangle_{13}+|\Psi^{-}\rangle_{13})\,|\Phi^{-}\rangle_{24}\nonumber\\&+&
(|\Psi^{+}\rangle_{13}-|\Phi^{-}\rangle_{13})\,|\Psi^{+}\rangle_{24}
-(|\Psi^{+}\rangle_{13}\nonumber\\&-&|\Psi^{-}\rangle_{13})\,
|\Psi^{-}\rangle_{24}]. \label{q4bellbasis}
\end{eqnarray}
From (4)  and (\ref{q4bellbasis}), it is
clear that one cannot generate key even if their basis matches.
Therefore, we have shown that although the four-qubit state
$|S_{1}\rangle$ has more entanglement than $|GHZ\rangle$ state but
the former cannot be used in the QKD protocol while the latter
can. Similarly we can also see that the state $|S_{2}\rangle$
has more entanglement in bipartite partitions, but it is not useful
for the QKD protocol. We note that the protocol will succeed for
the $|W\rangle$ and $|\Omega\rangle$ states. It would appear that,
like teleportation, only for specific values of the entropy of
the subsystems, our QKD protocol will succeed. Therefore
the resource state must have certain structure. One should be able to write
the state as $\sum_{k=1} a_{i} \ket{\varphi_{i}}  \ket{\psi_{i}} $,
where $\{ \ket{\varphi_{i}} \}$ and   $\{ \ket{\psi_{i}} \}$ are
orthonormal. The $|GHZ\rangle$,  $|W\rangle$ and $|\Omega\rangle$
can be written in this form, but $|S_{1}\rangle$ and $|S_{2}\rangle$
states cannot be. \\

\section{Conclusion}

Multipartite states allow  many variations of the communication
protocols that were introduced for bipartite states. We have
considered a number of different genuine quadripartite entangled
states as quantum resources for {\em exact} teleportation,
superdense coding and quantum key distribution protocols. With a
multipartite state as a resource, one can consider possibilities
of teleporting multiple qubit states with different number of
terms. We find that in such scenarios the phenomenon of ``less is
more'' may occur. It means that depending upon the value of
entropy, the resource state is either suitable or not suitable for
the teleportation of an unknown state. To understand this we have
obtained a condition that a resource state must satisfy for the
protocol to succeed. In particular, we find that to teleport a
$m$-term $n$-qubit state, a subsystem of the resource state must
have entropy ${\rm log}_{2} m$. So to be able to teleport a
two-term two-qubit state, a subsystem need to have the entropy as
one, but for a most general two-qubit teleportation, the required
entropy is two. Therefore, a four-qubit GHZ state can be used to
teleport a two-term two-qubit state but not the most general
two-qubit state.

For the {\em exact} superdense coding, the capacity also seems to
depend on the entropy of the subsystem on which Alice applies
unitary transformations to encode the classical information.
 We also have a number of
situations where one can encode more classical information by
applying unitary transformations on a subsystem with smaller
entropy. This again suggests that sometime ``less is more''.
Furthermore, for some specific distribution of qubits, sometimes
one can carry out superdense coding, but not the teleportation. It
also appears that if Alice's subsystem has certain entropy then
the superdense coding is possible. In particular for the maximal
superdense coding, the  Alice's subsystem should have maximal
entropy, allowed for her number of qubits. We also find the
concept of ``less is more'' holds for the quantum key distribution
schemes. In general, it appears that the
phenomenon of ``less is more'' is normal in the case of
communications protocols involving multipartite state. We note
that even when a resource state cannot be used to carry out
a protocol deterministically, it can be used for probabilistic
implementation of the protocol. However, it is always better to
carry out a protocol exactly, with less entangled state, than carry
it out probabilistically with more entangled state.

\end{document}